\documentclass[12pt, draftclsnofoot, onecolumn]{IEEEtran}

\ifCLASSINFOpdf
\else
   \usepackage[dvips]{graphicx}
\fi
\usepackage{url}

\usepackage{graphicx}
\usepackage{amsmath}
\usepackage{pgfplots}
\pgfplotsset{compat=1.18}

\usepackage{amssymb}
\usepackage{blkarray}
\usepackage{stmaryrd}
\usepackage[acronym]{glossaries}
\usepackage{cite}
\usepackage{algorithm}
\usepackage{algpseudocode}

\newcommand{\rev}[1]{#1}

\newcommand{\bdris}{\gls{bdris}}

\newcommand{\storm}{\gls{storm}}
\newcommand{\stars}{\gls{star}}

\makeglossaries

\newacronym{bs}{BS}{base station}
\newacronym{ris}{RIS}{reconfigurable intelligent surface}
\newacronym{ue}{UE}{user equipment}
\newacronym{bdris}{BD-RIS}{beyond diagonal reconfigurable intelligent surface}
\newacronym{ls}{LS}{least squares}
\newacronym{csi}{CSI}{channel state information}
\newacronym{omp}{OMP}{orthogonal matching pursuit}
\newacronym{dft}{DFT}{discrete Fourier transform}
\newacronym{aod}{AoD}{angle-of-departure}
\newacronym{aoa}{AoA}{angle-of-arrival}
\newacronym{svd}{SVD}{singular value decomposition}
\newacronym{mmwave}{mmWave}{millimeter wave}
\newacronym{subthz}{sub-THz}{sub-terahertz}
\newacronym{cs}{CS}{compressive sensing}
\newacronym{nmse}{NMSE}{normalized mean squared error}
\newacronym{snr}{SNR}{signal-to-noise ratio}
\newacronym{lso}{LSO}{LS Oracle}
\newacronym{storm}{STORM}{Sparse Tensor Orthogonal Recovery Method}
\newacronym{star}{STAR}{Sparse Tensor subspace-Aided Recovery}

\begin{document}

\title{Channel Estimation for Beyond Diagonal RIS Exploiting Core Tensor Sparsity}

\author{Daniel~Costa~Araújo and André~L.~F.~de~Almeida}

\maketitle

\begin{abstract}
\rev{\Glspl{bdris}} enhance wave manipulation through inter-element couplings but pose significant channel estimation challenges due to cascaded channels and block-Kronecker structures. This paper proposes a compressive sensing framework exploiting the sparse Tucker decomposition of the measurement tensor and the Kronecker rank-one structure of channel components. Two algorithms are developed: \gls{storm}, which uses \gls{omp} for greedy support recovery, and \gls{star}, which leverages subspace-based projection to enhance robustness to noise. Both perform joint sparse support identification, followed by a Kronecker rank-one \rev{factorization} via \gls{svd} to recover the channel parameters. Simulations show that \gls{star} achieves oracle-assisted \gls{ls} performance at moderate-to-high \gls{snr} with significantly fewer measurements than baseline methods, enabling practical \gls{bdris} deployment in next-generation \gls{mmwave}/\gls{subthz} networks.
\end{abstract}

\begin{IEEEkeywords}
   \gls{bdris}, \gls{cs}, \gls{mmwave}, \gls{ris}
\end{IEEEkeywords}

\IEEEpeerreviewmaketitle

\section{Introduction}

\IEEEPARstart{R}{econfigurable} intelligent surfaces (\gls{ris}s) have emerged as a transformative technology for sixth-generation (6G) wireless networks, enabling programmable control of electromagnetic wave propagation through planar arrays of low-cost passive elements~\cite{Gong2020_Survey, Wu2021_Tutorial, Liu2021_Principles, Shen2025_Tutorial}.
By dynamically adjusting the phase shifts induced by individual surface elements, conventional \gls{ris}s can perform passive beamforming, enhance coverage, and improve spectral efficiency without requiring active radio-frequency chains~\cite{Huang2019_EnergyEfficiency, Wu2020_DiscretePhase, 11011681}.
However, the diagonal scattering matrix structure of conventional \gls{ris}s, where each element operates independently, fundamentally constrains their wave manipulation flexibility~\cite{Li2024_RIS20}.

\rev{\Glspl{bdris}} overcome this limitation by introducing inter-element connections through reconfigurable impedance networks, yielding scattering matrices with non-zero off-diagonal entries~\cite{Li2024_RIS20, Li2023_MultiSector}.
This architectural innovation enables waves absorbed by one element to propagate into neighboring elements, thereby expanding the degrees of freedom for channel engineering~\cite{Nerini2024_GraphTheory, Li2024_MutualCoupling}.
The group-connected architecture organizes elements into groups with tunable intra-group couplings, offering a practical trade-off between performance and hardware complexity~\cite{Li2024_PhysicsCompliant}.
These advances translate into performance improvements, including multi-sector operation for full-space coverage~\cite{Li2023_MultiSector} and superior channel gain~\cite{Li2024_MutualCoupling, Li2025_Lossy}.

Despite these promising advantages, acquiring accurate \gls{csi} for \gls{bdris}-assisted systems is fundamentally more challenging than for conventional \gls{ris} due to several factors:
(i)~the passive nature of \gls{ris} elements precludes active pilot processing, necessitating estimation of cascaded transmitter-\gls{ris}-receiver channels~\cite{Zheng2022_ChannelEstSurvey, Swindlehurst2022_Framework,AraujodeAlmeida2020,AraujodeAlmeida2021};
(ii)~the inter-element coupling inherent in \gls{bdris} architectures introduces additional structural constraints that must be leveraged for efficient estimation~\cite{10587164, 10942770};
and (iii)~the large number of \gls{ris} elements, particularly in \gls{mmwave} and \gls{subthz} deployments, leads to prohibitive training overhead if not properly addressed~\cite{almeida2025channel}.

Recent research has proposed various approaches to tackle these challenges.
\Gls{ls}-based methods with optimized pilot sequences minimize estimation error under \gls{bdris} architectural constraints but require substantial training overhead~\cite{10587164}.
Decoupled estimation techniques exploit the block structure of \gls{bdris} scattering matrices, reducing problem dimensionality at the cost of potential information loss~\cite{10942770}.
Semi-blind methods based on PARATUCK tensor decomposition jointly estimate channels and data symbols, avoiding dedicated pilot overhead but with increased algorithmic complexity~\cite{10942666}.
More recently, block Tucker decomposition has been applied to \gls{bdris} channel estimation, yielding performance gains over matrix-based \gls{ls} methods due to its inherent noise-rejection properties~\cite {almeida2025channel}.

Motivated by the \rev{limited-scattering} nature of high-frequency propagation, \rev{typical of \gls{mmwave}/\gls{subthz} deployments where a \gls{bdris} restores blocked \gls{bs}-\gls{ue} coverage,} \gls{cs}-based methods have demonstrated significant reductions in training overhead for conventional \gls{ris} systems by exploiting channel sparsity in angular domains.
For instance, the TRICE framework combines tensor decomposition with ESPRIT-based direction estimation to efficiently recover sparse channels~\cite{9354904}.
Other techniques such as \gls{omp} leverage common support across users and subcarriers to recover sparse channel representations with reduced pilot sequences~\cite{Abdallah2023_RIS_CS}, while deep learning-enhanced \gls{cs} methods further improve recovery accuracy~\cite{Soumya2025_DeepCS}.
Nevertheless, direct application of existing \gls{cs} methods designed for conventional \gls{ris} to \gls{bdris} fails because the block-diagonal scattering structure introduces a unique sparsity pattern in the core tensor that is not captured by standard sparse recovery algorithms.

This paper proposes two novel \gls{cs}-based algorithms for \gls{bdris} channel estimation that exploit core tensor sparsity.
Both algorithms operate in two stages: first, a sparse support recovery stage identifies active propagation paths using Fourier dictionary matrices; second, a channel parameter estimation stage exploits the Kronecker rank-one structure of each support component through \gls{svd} and subspace projection to refine the estimates.
Our contributions are:
\begin{itemize}
\item \rev{We propose a \gls{bdris} channel-estimation scheme that jointly compresses the transmitter, receiver, and \gls{bdris} training dimensions, operating in a regime in which full-dimensional tensor estimators are not identifiable.}
\item \rev{We formulate \gls{bdris} channel estimation as a sparse core-tensor recovery problem and instantiate it via \gls{storm} (greedy \gls{omp}) and \gls{star} (subspace MUSIC).}
\item Through extensive simulations, we demonstrate that \gls{star} achieves oracle performance at high \gls{snr} while requiring significantly fewer training frames than conventional methods, with computational complexity lower than the TRICE framework~\cite{9354904} adapted to \gls{bdris}.
\end{itemize}

\section{System Model}

We consider a \gls{ris}-assisted system comprising a \gls{bs} with $N$ antennas, a \gls{ris} with $K$ elements, and a \gls{ue} with $M$ antennas. The \gls{ris} adopts a \gls{bdris} architecture with $Q$ groups of $\bar{K}$ elements each, where $K = \bar{K} Q$. During training, beam matrices $\mathbf{W}_{\text{tx}} \in \mathbb{C}^{N \times N_{\text{tx}}}$ and $\mathbf{W}_{\text{rx}} \in \mathbb{C}^{M \times M_{\text{rx}}}$ are employed\rev{, with $N_{\text{tx}}\!\ll\!N$ and $M_{\text{rx}}\!\ll\!M$, a regime under which the full-dimensional tensor estimator of~\cite{almeida2025channel} is not identifiable}. Let $\mathbf{G} \in \mathbb{C}^{N \times K}$ and $\mathbf{H} \in \mathbb{C}^{M \times K}$ denote the \gls{bs}-\gls{ris} and \gls{ue}-\gls{ris} channels. For the $\ell$-th training frame, each group $q$ has scattering matrix $\mathbf{W}_{\text{ris}}^{(q,\ell)} \in \mathbb{C}^{\bar{K} \times \bar{K}}$, forming the block diagonal matrix
\begin{equation}
\label{block_diagonal}
\mathbf{B}^{(\ell)} = \text{blkdiag}\left[\mathbf{W}_{\text{ris}}^{(1,\ell)}, \ldots, \mathbf{W}_{\text{ris}}^{(Q,\ell)}\right].
\end{equation}

\rev{Consider $\mathbf{H}_q, \mathbf{G}_q$ as the column sub-matrices of $\mathbf{H}, \mathbf{G}$ for the $\bar{K}$ elements of group $q$.} \rev{With the processed channels $\bar{\mathbf{H}}_q = \mathbf{W}_{\text{rx}} \mathbf{H}_q \in \mathbb{C}^{M_{\text{rx}} \times \bar{K}}$ and $\bar{\mathbf{G}}_q = \mathbf{W}_{\text{tx}} \mathbf{G}_q \in \mathbb{C}^{N_{\text{tx}} \times \bar{K}}$, the signal during frame $\ell$ is}
\begin{equation}
\label{system_model}
\mathbf{Y}^{(\ell)} = \sum_{q=1}^{Q} \bar{\mathbf{H}}_q \mathbf{W}_{\text{ris}}^{(q,\ell)} \bar{\mathbf{G}}_q^T  + \mathbf{N}^{(\ell)},
\end{equation}
where $\mathbf{N}^{(\ell)}$ is the processed noise matrix. The channels are \rev{modeled} as $\mathbf{H} = \mathbf{U}_H \boldsymbol{\Lambda}_H \mathbf{V}_H^H$ and $\mathbf{G} = \mathbf{U}_G \boldsymbol{\Lambda}_G \mathbf{V}_G^H$, where $\mathbf{U}_H$, $\mathbf{U}_G$ contain steering vectors at \gls{ue} and \gls{bs}, $\mathbf{V}_H$, $\mathbf{V}_G$ at the \gls{ris}, and $\boldsymbol{\Lambda}_H$, $\boldsymbol{\Lambda}_G$ are diagonal path gain matrices with $P$ paths each\rev{, set to $P_G=P_H=P$ for notational simplicity, with the framework extending to $P_G\neq P_H$ via sparsity $S=P_GP_H$}. \rev{Owing to limited \gls{mmwave}/\gls{subthz} scattering, $P$ is small and each link is angularly sparse.} We introduce the block Kronecker product operator $\mathbf{A} |\otimes| \mathbf{B} = \left[\mathbf{A}_1 \otimes \mathbf{B}_1, \ldots, \mathbf{A}_Q \otimes \mathbf{B}_Q\right]$ and define $\bar{\mathbf{V}} = \mathbf{V}_G^H |\otimes| \mathbf{V}_H^H$. By \rev{vectorizing}~\eqref{system_model} and concatenating all $K_{\text{ris}}$ training frames, while collecting the \rev{vectorized} \gls{ris} configurations as $\mathbf{w}_{\text{ris}}^{(\ell)} = \left[\text{vec}(\mathbf{W}_{\text{ris}}^{(1,\ell)})^T, \ldots, \text{vec}(\mathbf{W}_{\text{ris}}^{(Q,\ell)})^T\right]^T$ into $\mathbf{W}_{\text{ris}} = \left[\mathbf{w}_{\text{ris}}^{(1)}, \ldots, \mathbf{w}_{\text{ris}}^{(K_{\text{ris}})}\right] \in \mathbb{C}^{Q\bar{K}^2 \times K_{\text{ris}}}$, the noise-free received signal across all frames becomes \rev{$\mathbf{X} = \left(\mathbf{W}_{\text{tx}} \mathbf{U}_G \otimes \mathbf{W}_{\text{rx}} \mathbf{U}_H\right) \left(\boldsymbol{\Lambda}_G \otimes \boldsymbol{\Lambda}_H\right) \bar{\mathbf{V}} \mathbf{W}_{\text{ris}}$}, with $\mathbf{X} = \left[\text{vec}(\mathbf{X}^{(1)}), \ldots, \text{vec}(\mathbf{X}^{(K_{\text{ris}})})\right]$. Taking the transpose reveals a Tucker-3 structure:
\begin{equation}
\label{tucker3_decomposition}
\mathbf{X}^T = \mathbf{A}_3 \mathbf{D}_{(3)} \left(\mathbf{A}_2 \otimes \mathbf{A}_1\right)^T,
\end{equation}
where $\mathbf{A}_1 = \mathbf{W}_{\text{tx}} \mathbf{U}_G \in \mathbb{C}^{N_{\text{tx}} \times P}$, $\mathbf{A}_2 = \mathbf{W}_{\text{rx}} \mathbf{U}_H \in \mathbb{C}^{M_{\text{rx}} \times P}$, $\mathbf{A}_3 = \mathbf{W}_{\text{ris}}^T \bar{\mathbf{V}}^T \in \mathbb{C}^{K_{\text{ris}} \times P^2}$, and $\mathbf{D}_{(3)} = (\boldsymbol{\Lambda}_G \otimes \boldsymbol{\Lambda}_H)^T$ is the mode-3 unfolding of the core tensor $\mathcal{D}$. The tensor model is $\mathcal{Y} = \mathcal{X} + \mathcal{N}$, where $\mathcal{Y}, \mathcal{X}, \mathcal{N} \in \mathbb{C}^{N_{\text{tx}} \times M_{\text{rx}} \times K_{\text{ris}}}$.

A conventional \gls{ls} approach requires $K_{\text{ris}} > Q\bar{K}^2$ frames for identifiability, which becomes prohibitive as $\bar{K}$ grows~\cite{almeida2025channel}. However, in \gls{mmwave}/\gls{subthz} scenarios, $P \ll K$, so $\mathcal{D}$ is sparse and can be recovered with far fewer measurements through \gls{cs} techniques.

\section{Proposed Channel Estimation}

We propose a \gls{cs}-based channel estimation algorithm that exploits the sparsity structure of the core tensor $\mathcal{D}$ in \eqref{tucker3_decomposition}. The proposed approach consists of two main stages: sparse support recovery and channel parameter estimation.

\subsection{Stage 1: Sparse Support Recovery}

We construct \gls{dft} dictionary matrices $\mathbf{F}_1 \in \mathbb{C}^{N \times N_{\text{grid}}}$, $\mathbf{F}_2 \in \mathbb{C}^{M \times N_{\text{grid}}}$, \rev{$\mathbf{F}_G, \mathbf{F}_H \in \mathbb{C}^{K\times N_{\text{grid}}}$}, and $\mathbf{F}_3 = (\mathbf{F}_G^H |\otimes| \mathbf{F}_H^H) \in \mathbb{C}^{Q\bar{K}^2 \times N_{\text{grid}}^2}$\rev{, all discretizing the respective array steering vectors on $N_{\text{grid}}$ bins}. Defining $\bar{\mathbf{A}}_1 = \mathbf{W}_{\text{tx}} \mathbf{F}_1$, $\bar{\mathbf{A}}_2 = \mathbf{W}_{\text{rx}} \mathbf{F}_2$, and $\bar{\mathbf{A}}_3 = \mathbf{W}_{\text{ris}}^T \mathbf{F}_3$, the \gls{cs} problem is
\begin{equation}
\label{cs_problem}
\rev{\mathbf{X}_{(3)}^T = \bar{\mathbf{A}}_3\,\boldsymbol{\theta} + \mathbf{N}_{(3)}^T,}
\end{equation}
where $\boldsymbol{\theta} \in \mathbb{C}^{N_{\text{grid}}^2 \times N_{\text{tx}}M_{\text{rx}}}$ is row-sparse with $P^2$ non-zero rows \rev{indexed by the active $(p_G,p_H)$ pairs, each of the form $d_s(\bar{\mathbf{A}}_2 \otimes \bar{\mathbf{A}}_1)(:,s)^T$ so that the TX/RX dictionaries are exposed only in the Stage~2 reshape}. We propose two strategies:

\textit{\gls{storm} (OMP-based):} Applies \gls{omp} to~\eqref{cs_problem} using $\bar{\mathbf{A}}_3$ as dictionary, iteratively selecting atoms with maximum residual correlation over $S = |\mathcal{S}|$ iterations.

\textit{\gls{star} (Subspace-based):} Computes the \gls{svd} of $\mathbf{X}_{(3)}^T$ and identifies the support via the MUSIC-like spectrum
\begin{equation}
\label{music_stage1}
\hat{s} = \arg\min_{k \in \{1,\ldots,N_{\text{grid}}^2\}} \|\bar{\mathbf{A}}_3(:,k)^H \mathbf{U}_{\text{null}}\|_F^2,
\end{equation}
selecting the $S$ indices with the smallest noise-subspace projection. For both methods, once $\mathcal{S}$ is determined, the coefficients are estimated via $\hat{\boldsymbol{\theta}}_{\mathcal{S}} = \bar{\mathbf{A}}_{3,\mathcal{S}}^{\dagger} \mathbf{X}_{(3)}^T$.

\subsection{Stage 2: Channel Parameter Estimation}

\rev{For each support index $s \in \mathcal{S}$, the row $\hat{\boldsymbol{\theta}}_{s,:}$ of length $N_{\text{tx}}M_{\text{rx}}$ is reshaped column-wise into $\mathbf{M}_s = \mathrm{unvec}_{N_{\text{tx}}\times M_{\text{rx}}}(\hat{\boldsymbol{\theta}}_{s,:}^T) \in \mathbb{C}^{N_{\text{tx}} \times M_{\text{rx}}}$, which is rank-one because each non-zero row of $\boldsymbol{\theta}$ corresponds to a single path pair $(p_G,p_H)$:}
\begin{equation}
\label{rank_one_structure}
\mathbf{M}_s = \mathbf{A}_1(:,p_G) \mathbf{A}_2(:,p_H)^T.
\end{equation}
\rev{Since $s$ encodes $(p_G,p_H)$, no path-pair matching is needed, and off-grid leakage is mitigated by oversampled dictionaries.}

Performing \gls{svd} on $\mathbf{M}_s = \mathbf{U}_s \boldsymbol{\Sigma}_s \mathbf{V}_s^H$ and defining noise subspaces $\mathbf{U}_{\text{null}}$ and $\mathbf{V}_{\text{null}}$ from the trailing $(N_{\text{tx}}-1)$ and $(M_{\text{rx}}-1)$ singular vectors of $\mathbf{U}_s$ and $\mathbf{V}_s$, respectively, we estimate the path indices via
\begin{equation}
\label{subspace_search}
\rev{\begin{split}
\hat{p}_{\text{tx}} &= \arg\min_{i} \|\bar{\mathbf{A}}_1(:,i)^H \mathbf{U}_{\text{null}}\|_F^2, \\
\hat{p}_{\text{rx}} &= \arg\min_{j} \|\bar{\mathbf{A}}_2(:,j)^T \mathbf{V}_{\text{null}}\|_F^2,
\end{split}}
\end{equation}
and compute $\hat{d}_{s} = \bar{\mathbf{A}}_1(:,\hat{p}_{\text{tx}})^H \mathbf{M}_s \bar{\mathbf{A}}_2(:,\hat{p}_{\text{rx}})^*$. \rev{After processing all $s \in \mathcal{S}$, the entries $\hat{d}_s$ form $\hat{\mathbf{D}}_{(3),\mathcal{S}} = \text{diag}(\hat{d}_1, \ldots, \hat{d}_{|\mathcal{S}|})$ and the composite channel is}
\begin{equation}
\label{channel_estimate}
\widehat{\bar{\mathbf{G}} |\otimes| \bar{\mathbf{H}}} = \mathbf{F}_{3,\mathcal{S}} \hat{\mathbf{D}}_{(3),\mathcal{S}} \left(\mathbf{F}_{2,\mathcal{S}} \otimes \mathbf{F}_{1,\mathcal{S}}\right)^T.
\end{equation}
Algorithms~\ref{alg:storm} and~\ref{alg:star} \rev{summarize} the complete procedures for \gls{storm} and \gls{star}, highlighting the Stage~1 difference that governs the performance trade-off between the two methods.

\begin{algorithm}[t]
\caption{\Gls{storm}: OMP-based sparse tensor recovery}
\label{alg:storm}
\begin{algorithmic}[1]
\Require $\mathbf{X}_{(3)}^T$, dictionaries $\bar{\mathbf{A}}_1,\bar{\mathbf{A}}_2,\bar{\mathbf{A}}_3$, sparsity $S$
\Ensure Estimated composite channel $\widehat{\bar{\mathbf{G}} |\otimes| \bar{\mathbf{H}}}$
\Statex \textit{Stage 1: OMP-based support recovery}
\State $\mathcal{S} \gets \emptyset$,\quad $\mathbf{R} \gets \mathbf{X}_{(3)}^T$
\For{$t = 1, \ldots, S$}
  \State $k^* \gets \arg\max_{k}\,\|\bar{\mathbf{A}}_3(:,k)^H \mathbf{R}\|_F$ \Comment{correlation}
  \State $\mathcal{S} \gets \mathcal{S} \cup \{k^*\}$
  \State $\hat{\boldsymbol{\theta}}_{\mathcal{S}} \gets \bar{\mathbf{A}}_{3,\mathcal{S}}^{\dagger}\,\mathbf{X}_{(3)}^T$ \Comment{LS update}
  \State $\mathbf{R} \gets \mathbf{X}_{(3)}^T - \bar{\mathbf{A}}_{3,\mathcal{S}}\hat{\boldsymbol{\theta}}_{\mathcal{S}}$ \Comment{residual update}
\EndFor
\Statex \textit{Stage 2: Kronecker rank-one \rev{factorization}}
\For{each $s \in \mathcal{S}$}
  \State Reshape $\hat{\boldsymbol{\theta}}_{\mathcal{S}(s),:}^T$ into $\mathbf{M}_s \in \mathbb{C}^{N_{\text{tx}} \times M_{\text{rx}}}$
  \State $[\mathbf{U}_s, \boldsymbol{\Sigma}_s, \mathbf{V}_s] \gets \mathrm{SVD}(\mathbf{M}_s)$
  \State $\mathbf{U}_{\text{null}}^{(s)} \gets$ trailing $(N_{\text{tx}}-1)$ cols.\ of $\mathbf{U}_s$
  \State $\mathbf{V}_{\text{null}}^{(s)} \gets$ trailing $(M_{\text{rx}}-1)$ cols.\ of $\mathbf{V}_s$
  \State Estimate $\hat{p}_{\text{tx}}, \hat{p}_{\text{rx}}$ via \eqref{subspace_search}
  \State $\hat{d}_s \gets \bar{\mathbf{A}}_1(:,\hat{p}_{\text{tx}})^H\mathbf{M}_s\,\bar{\mathbf{A}}_2(:,\hat{p}_{\text{rx}})^*$
\EndFor
\State $\hat{\mathbf{D}}_{(3),\mathcal{S}} \gets \mathrm{diag}(\hat{d}_1,\ldots,\hat{d}_{|\mathcal{S}|})$
\State \Return $\mathbf{F}_{3,\mathcal{S}}\,\hat{\mathbf{D}}_{(3),\mathcal{S}}\,(\mathbf{F}_{2,\mathcal{S}} \otimes \mathbf{F}_{1,\mathcal{S}})^T$
\end{algorithmic}
\end{algorithm}

\begin{algorithm}[t]
\caption{\Gls{star}: Subspace-based sparse tensor recovery}
\label{alg:star}
\begin{algorithmic}[1]
\Require $\mathbf{X}_{(3)}^T$, dictionaries $\bar{\mathbf{A}}_1,\bar{\mathbf{A}}_2,\bar{\mathbf{A}}_3$, sparsity $S$
\Ensure Estimated composite channel $\widehat{\bar{\mathbf{G}} |\otimes| \bar{\mathbf{H}}}$
\Statex \textit{Stage 1: Subspace-based support recovery}
\State $[\mathbf{U}, \boldsymbol{\Sigma}, \mathbf{V}] \gets \mathrm{SVD}\!\left(\mathbf{X}_{(3)}^T\right)$
\State $\mathbf{U}_{\text{null}} \gets$ trailing $(K_{\text{ris}} - S)$ columns of $\mathbf{U}$
\For{$k = 1, \ldots, N_{\text{grid}}^2$}
  \State $f(k) \gets \left\|\bar{\mathbf{A}}_3(:,k)^H\mathbf{U}_{\text{null}}\right\|_F^2$ \Comment{MUSIC spectrum}
\EndFor
\State $\mathcal{S} \gets$ indices of the $S$ smallest $f(k)$
\State $\hat{\boldsymbol{\theta}}_{\mathcal{S}} \gets \bar{\mathbf{A}}_{3,\mathcal{S}}^{\dagger}\,\mathbf{X}_{(3)}^T$
\Statex \textit{Stage 2: Execute Stage~2 of Algorithm~\ref{alg:storm} (Steps~8--14)}
\end{algorithmic}
\end{algorithm}

\begin{table}[!t]
\centering
\caption{Computational Complexity Comparison}
\label{tab:complexity}
\begin{tabular}{lcc}
\hline
\textbf{Method} & \textbf{Stage 1} & \textbf{Stage 2} \\
\hline
\gls{storm}& $\mathcal{O}(S K_{\text{ris}} N_{\text{grid}}^2 N_{\text{tx}}M_{\text{rx}})$ & $\mathcal{O}(S N_{\text{grid}}(N_{\text{tx}}^2 + M_{\text{rx}}^2))$ \\[0.5ex]
\gls{star}  & \rev{$\mathcal{O}(K_{\text{ris}}^{2}(N_{\text{grid}}^2+N_{\text{tx}}M_{\text{rx}}))$} & $\mathcal{O}(S N_{\text{grid}}(N_{\text{tx}}^2 + M_{\text{rx}}^2))$ \\[0.5ex]
\rev{Vectorised CS}  & $\mathcal{O}(S K_{\text{ris}} N_{\text{tx}}M_{\text{rx}} N_{\text{grid}}^4)$ & --- \\[0.5ex]
TRICE \cite{9354904} & $\mathcal{O}(K_{\text{ris}}^2 Q\bar{K}^2 N_{\text{tx}}M_{\text{rx}})$ & $\mathcal{O}(K_{\text{ris}} Q\bar{K}^2)$ \\[0.5ex]
\hline
\end{tabular}
\end{table}

\subsection{Computational Complexity Analysis}

We \rev{analyze} the floating-point complexity of \gls{storm} and \gls{star}. Let $S=|\mathcal{S}|$, $N_{\text{grid}}$ be the angle-grid size, and $P$ the number of paths per channel.

\rev{The \gls{omp}-based support recovery in \gls{storm} runs $S$ iterations, each dominated by the correlation $\bar{\mathbf{A}}_3^H \mathbf{R}$ at cost $\mathcal{O}(K_{\text{ris}} N_{\text{grid}}^2 N_{\text{tx}} M_{\text{rx}})$, yielding a Stage~1 complexity $\mathcal{C}_{\text{Stage 1}}^{\text{STORM}}=\mathcal{O}(S K_{\text{ris}} N_{\text{grid}}^2 N_{\text{tx}} M_{\text{rx}})$. For \gls{star}, the Stage~1 cost is instead dominated by a single \gls{svd} of the mode-3 unfolding and by a one-shot projection of the $N_{\text{grid}}^{2}$ dictionary atoms onto the noise subspace, with total complexity $\mathcal{C}_{\text{Stage 1}}^{\text{STAR}}=\mathcal{O}\bigl(K_{\text{ris}}^{2}(N_{\text{grid}}^{2}+N_{\text{tx}} M_{\text{rx}})\bigr)$, independent of the support size $S$.}

Stage 2 is identical for both \gls{storm} and \gls{star}. For each of the $S$ identified support indices, Stage 2 performs the following operations: (i) computing the pseudoinverse $\bar{\mathbf{A}}_{3,\mathcal{S}}^{\dagger}$, which has complexity $\mathcal{O}\left(\min(K_{\text{ris}}, S)^2 \max(K_{\text{ris}}, S)\right)$ when computed once for the entire support set; (ii) reshaping and \gls{svd} of $\mathbf{M}_s \in \mathbb{C}^{N_{\text{tx}} \times M_{\text{rx}}}$, with complexity $\mathcal{O}\left(N_{\text{tx}} M_{\text{rx}} \min(N_{\text{tx}}, M_{\text{rx}})\right)$ per support index; and (iii) noise subspace search over the dictionary atoms as in \eqref{subspace_search}, which projects all columns of $\bar{\mathbf{A}}_1$ and $\bar{\mathbf{A}}_2$ onto the noise subspaces $\mathbf{V}_{\text{null}} \in \mathbb{C}^{N_{\text{tx}} \times (N_{\text{tx}}-1)}$ and $\mathbf{U}_{\text{null}} \in \mathbb{C}^{M_{\text{rx}} \times (M_{\text{rx}}-1)}$, requiring $\mathcal{O}\left(N_{\text{grid}}(N_{\text{tx}}^2 + M_{\text{rx}}^2)\right)$ operations per index. Aggregating over all $S$ support indices, the total Stage 2 complexity is $\mathcal{O}\left(S N_{\text{grid}}(N_{\text{tx}}^2 + M_{\text{rx}}^2)\right)$.

\rev{Since $S \ll N_{\text{grid}}^2$ and $K_{\text{ris}} \ll Q\bar{K}^2$, the total complexity is dominated by Stage~1 for large $N_{\text{grid}}$. Both methods achieve lower complexity than TRICE~\cite{9354904}, $\mathcal{O}(K_{\text{ris}}^2 Q\bar{K}^2 N_{\text{tx}}M_{\text{rx}})$, by exploiting the Kronecker rank-one structure. The Stage~1 difference drives the trade-off: \gls{storm} uses greedy \gls{omp}, fast at high \gls{snr}, while \gls{star} uses a MUSIC-like subspace search, more robust at low \gls{snr}. The sparsity $S$ enters Stage~1 only for \gls{storm}, which scales linearly with $S$, whereas \gls{star} is sparsity-independent in Stage~1. The shared Stage~2 grows linearly with $S$ for both.}

Table~\ref{tab:complexity} \rev{summarizes} the computational complexity comparison among different methods. \rev{As a} matter of comparison, we \rev{analyse} the \rev{vectorised} \gls{cs} approach, which applies \gls{omp} directly to the flattened tensor measurement vector of dimension $K_{\text{ris}} N_{\text{tx}}M_{\text{rx}}$, constructing a dictionary via the Khatri-Rao product $\text{KR}(\mathbf{A}_1 \otimes \mathbf{A}_2, \mathbf{A}_3)$ with $N_{\text{grid}}^4$ atoms, covering all possible angle combinations across the three tensor modes. This single-stage \gls{omp} approach incurs prohibitively high complexity $\mathcal{O}\left(S K_{\text{ris}} N_{\text{tx}}M_{\text{rx}} N_{\text{grid}}^4\right)$ that scales quartically with the grid size and does not benefit from the two-stage decomposition. In contrast, by exploiting the tensor structure and Kronecker \rev{factorization}, \gls{storm} and \gls{star} reduce the dictionary size from $N_{\text{grid}}^4$ to $N_{\text{grid}}^2$, achieving a quadratic reduction in complexity. The TRICE framework, while also exploiting tensor structure, incurs a higher Stage 1 cost because it performs full tensor decomposition without exploiting core tensor sparsity. As shown in the table, \gls{storm} and \gls{star} share identical Stage 2 complexity but differ in their Stage 1 support recovery mechanisms, with \gls{star} providing enhanced noise robustness at the expense of a slightly higher computational cost during the support estimation phase.

\section{Numerical Results}
\label{sec:num_results}

We evaluate the proposed scheme through Monte Carlo simulations, reporting the \gls{nmse} versus \gls{snr}, the \gls{nmse} versus the fraction of \gls{ris} measurements relative to $Q\bar{K}^{2}$, the \gls{nmse} versus the number of paths, and the processing time versus $\bar{K}$. \rev{Default parameters are listed in Table~\ref{tab:sim_params}; $P$ is assumed known and estimable in practice via classical MDL/AIC rank tests on $\mathbf{X}_{(3)}$.} We compare the proposed \gls{storm} and \gls{star} with the TRICE framework~\cite{9354904} adapted to \gls{bdris} and the \gls{lso}, which knows the true support and serves as a lower bound. \rev{$\mathbf{W}_{\text{tx}}$, $\mathbf{W}_{\text{rx}}$ and the per-group $\mathbf{W}_{\text{ris}}^{(q,\ell)}$ are i.i.d.\ complex Gaussian, with $\mathbf{W}_{\text{ris}}^{(q,\ell)}$ assembled via~\eqref{block_diagonal} to enforce the group-connected structure. \gls{snr} is the per-element ratio $P_{s}/\sigma^{2}$ on $\mathcal{Y}=\mathcal{X}+\mathcal{N}$ of~\eqref{system_model}, with $\mathcal{N}\!\sim\!\mathcal{CN}(0,\sigma^{2}\mathbf{I})$, and the path gains are $\alpha_{p}\!\sim\!\mathcal{CN}(0,1/P)$. The TRICE baseline runs a joint-support \gls{omp} with sparsity $S=P^{2}$ over $\bar{\mathbf{A}}_{2}\otimes\bar{\mathbf{A}}_{1}$ on the mode-2 unfolding, then a per-column sparsity-one \gls{omp} over $\bar{\mathbf{A}}_{3}$. The \gls{lso} solves a Khatri-Rao LS fit on the true support.}

\begin{table}[!t]
\centering
\rev{\caption{Default Simulation Parameters}
\label{tab:sim_params}
\renewcommand{\arraystretch}{1.05}
\setlength{\tabcolsep}{5pt}
\footnotesize
\begin{tabular}{lcc}
\hline
\textbf{Parameter} & \textbf{Symbol} & \textbf{Value} \\
\hline
\gls{bs}/\gls{ue} antennas & $N=M$ & $32$ \\
\gls{bdris} elements / group size & $K$ / $\bar{K}$ & $64$ / $\{4,8\}$ \\
TX/RX training beams & $N_{\text{tx}}=M_{\text{rx}}$ & $16$ \\
Paths per channel & $P$ & $2$ \\
Angular grids ($2\times$ oversampling) & $N_{\text{grid}}, N_{\text{ris,grid}}$ & $2N$, $2K$ \\
\gls{snr} per element & \gls{snr} (dB) & $\{0,4,\ldots,20\}$ \\
Monte Carlo trials per point & $N_{\mathrm{MC}}$ & $1{,}000$ \\
\hline
\end{tabular}}
\end{table}

\begin{figure}[t]
  \centering
  \includegraphics[width=\columnwidth]{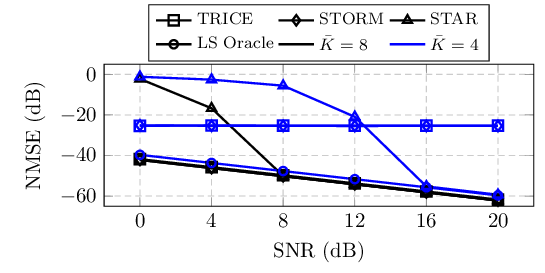}
  \caption{\gls{nmse} as a function of \gls{snr}. Fixed: $N=32$, $M=32$, $K=64$, $N_{\text{tx}}=16$, $M_{\text{rx}}=16$; \gls{ris} measurement size at 50\% of $Q\bar{K}^2$.}
  \label{fig:nmse_snr}
\end{figure}

Fig.~\ref{fig:nmse_snr} shows the \gls{nmse} versus \gls{snr}. The curves reflect the impact of $\bar{K}$: a larger $\bar{K}$ offers greater robustness at low \gls{snr}, with a smaller \gls{nmse} gap to the bound in that regime compared with $\bar{K}=4$. For $\bar{K}=8$, all algorithms converge to the \gls{lso} bound as the \gls{snr} increases, confirming that the measurement budget at 50\% compression is sufficient for accurate sparse recovery when the group size provides enough diversity. \rev{For $\bar{K}=4$, only the subspace-based \gls{star} attains the bound at high \gls{snr}, while the \gls{omp}-based TRICE~\cite{9354904} and \gls{storm} saturate at an \gls{snr}-independent floor. The smaller Kronecker dimension $\bar{K}^{2}=16$ raises atom coherence, causing \gls{omp} to misselect support atoms regardless of the noise level.}

\begin{figure}[t]
  \centering
  \includegraphics[width=\columnwidth]{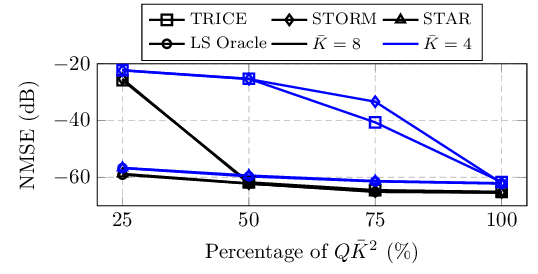}
  \caption{\gls{nmse} as a function of the percentage of \gls{ris} measurements relative to $Q\bar{K}^2$. Fixed: $N=32$, $M=32$, $K=64$, $N_{\text{tx}}=16$, $M_{\text{rx}}=16$.}
  \label{fig:nmse_ris_pct}
\end{figure}

Fig.~\ref{fig:nmse_ris_pct} plots the \gls{nmse} versus the fraction of $Q\bar{K}^2$ used as measurements\rev{, i.e., the training overhead normalized by the \gls{ls} threshold $Q\bar{K}^{2}$}. The \gls{lso} bound improves as the percentage increases due to the larger effective aperture of the measurement matrix. For $\bar{K}=8$, all algorithms converge to the bound as the measurement fraction grows, indicating that even moderate compression ratios (30--50\%) are sufficient for reliable sparse recovery. Notably, \gls{star} achieves near-oracle performance across both group sizes: for $\bar{K}=4$, \gls{star} tracks the \gls{lso} bound closely even under heavy compression, whereas \gls{storm} exhibits a larger gap. This demonstrates that the subspace-based approach of \gls{star} provides superior robustness to the reduced degrees of freedom in smaller group configurations compared to the greedy \gls{omp}-based strategy of \gls{storm}. From a practical standpoint, these results suggest that \gls{bdris} systems can operate with significantly fewer training frames than the $Q\bar{K}^2$ required by conventional \gls{ls} methods, freeing resources for data transmission. \rev{Formal bounds for~\eqref{cs_problem} follow from the coherence/spark of $\bar{\mathbf{A}}_{3}$~\cite{Caiafa2013_KroneckerCS}.}

\begin{figure}[t]
  \centering
  \includegraphics[width=\columnwidth]{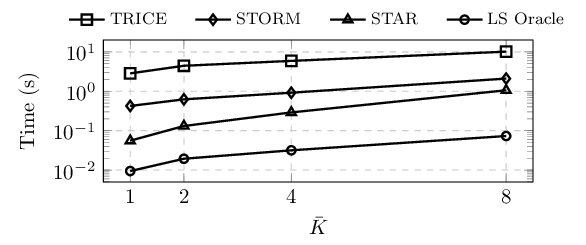}
  \vspace{-1.6ex}
  \caption{Processing time of the algorithms versus $\bar{K}$. Fixed: $N=32$, $M=32$, $K=64$, $N_{\text{tx}}=16$, $M_{\text{rx}}=16$; \gls{ris} measurement size at 50\% of $Q\bar{K}^2$.}
  \label{fig:time_nbar}
\end{figure}

\rev{Fig.~\ref{fig:time_nbar} shows processing time versus $\bar{K}$. \stars{} and \storm{} scale much more gently than TRICE, keeping larger group sizes practical and covering fully connected and hybrid \bdris{} topologies.}

\begin{figure}[t]
  \centering
  \includegraphics[width=\columnwidth]{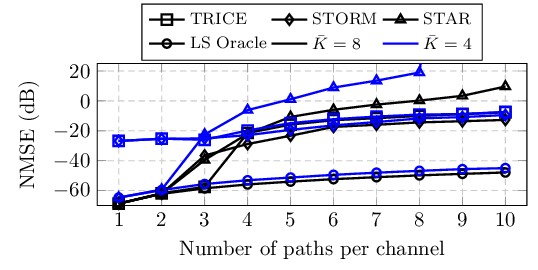}
  \caption{\gls{nmse} as a function of the number of paths per channel. Fixed: $N=32$, $M=32$, $K=64$, $N_{\text{tx}}=16$, $M_{\text{rx}}=16$; \gls{ris} measurement size at 50\% of $Q\bar{K}^2$.}
  \label{fig:nmse_n_paths}
\end{figure}

Fig.~\ref{fig:nmse_n_paths} shows the \gls{nmse} versus the number of paths per channel with a fixed number of measurements at 50\% of $Q\bar{K}^2$. As $P$ increases, all algorithms experience performance degradation because the effective sparsity level decreases; the measurement budget remains constant while the number of unknowns grows as $P^2$. This imbalance is particularly evident for \gls{star}, which exhibits a steeper degradation curve than \gls{storm}. This behavior arises because the subspace-based support identification in \gls{star} relies on a clear separation between the signal and noise subspaces, which degrades as the core tensor becomes less sparse. Nevertheless, for the typical path counts encountered in \gls{mmwave} and \gls{subthz} environments ($P \leq 4$), both proposed methods maintain \gls{nmse} values well below $-20$~dB, confirming their suitability for practical deployment. \rev{For denser-scattering scenarios, a non-sparse estimator such as the one in~\cite{almeida2025channel} is more appropriate.}

\section{Conclusion}

This paper addressed the channel estimation problem for \gls{bdris}-assisted wireless systems by formulating it as a sparse core tensor recovery problem that exploits the limited-scattering properties of \gls{mmwave} and \gls{subthz} channels. \rev{The two proposed algorithms, STORM and STAR, leverage Kronecker rank-one factorization to estimate sparse channels with reduced training overhead, and simulations confirm near-oracle \gls{nmse}, with STAR attaining the oracle bound at high \gls{snr} while STORM and TRICE saturate at the dictionary-coherence floor for small group sizes. Both methods retain a clear complexity advantage over TRICE adapted to \gls{bdris}, with the largest compression gains at $\bar{K}=8$. Future work targets wideband OFDM, hybrid group/fully-connected topologies, off-grid Stage-2 refinement, and direct-link extensions.}

\renewcommand\baselinestretch{1.0}
\bibliographystyle{IEEEtran}
\bibliography{references_v2}

\end{document}